\documentclass[twocolumn,aps]{revtex4}%
\usepackage{amsfonts}
\usepackage{amsmath}
\usepackage{amssymb}
\usepackage{graphicx}%
\begin{document}
\preprint{ }
\title[ ]{Stochastic method for accommodation of equilibrating basins in kinetic Monte
Carlo simulations}
\author{Clinton DeW. Van Siclen}
\email{clinton.vansiclen@inl.gov}
\affiliation{Idaho National Laboratory, Idaho Falls, Idaho 83415, USA}

\begin{abstract}
A computationally simple way to accommodate `basins'\ of trapping states in
standard kinetic Monte Carlo simulations is presented. By assuming the system
is effectively equilibrated in the basin, the residence time (time spent in
the basin before escape) and the probabilities for transition to states
outside the basin may be calculated. This is demonstrated for point defect
diffusion over a periodic grid of sites containing a complex basin.

\end{abstract}
\volumeyear{year}
\volumenumber{number}
\issuenumber{number}
\eid{identifier}
\published{2 February 2007}

\startpage{1}
\endpage{102}
\maketitle

\footnotetext[1]{Published: J. Phys.: Condens. Matter \textbf{19}, 072201
(2007).}The kinetic Monte Carlo (kMC) method is used to evolve atomistic
systems \textit{dynamically} from state to state over timescales much longer
than can be achieved in molecular dynamics simulations \cite{r1,r2}. The
method utilizes a catalog of state-to-state transition rates obtained from
atomistic (dynamic or static) calculations, to determine probabilistically a
sequence of states (and their residence times) that closely resembles the
actual system dynamics. The computational efficiency of the kMC method is due
to the neglect of details: the system is simply moved from one distinct state
to another, and the time clock is advanced accordingly. However, it may be
that the set of transition rates is such as to equilibrate the system in a
subset of mutually accessible states, from which escape is a very rare event.
This situation of course reduces the efficiency of the method greatly. Here we
present a simple means to accommodate such equilibrating basins in the
standard kMC approach for the case of defect diffusion in solids or on
surfaces. In fact the basin is regarded as just another accessible defect site
with a characteristic residence time. This is only possible when the defect is
considered to have equilibrated in the basin (that is, all sites in the basin
have been visited many times), so its entry and exit points are uncorrelated.
This treatment of equilibrating basins will be particularly useful for kMC
simulations of defect diffusion in nanocrystalline materials, where the
diffusion coefficients for the defect in the grain boundaries and the
crystalline grains may differ by many orders of magnitude \cite{r3}, and of
radiation damage in solids, where microstructure evolution (driven by defect
diffusion) over very long time scales is of interest.

In a kMC simulation of defect diffusion, the defect moves over a regular or
irregular grid of sites (representing the potential wells that can accommodate
the defect) according to probabilistic rules. The diffusion coefficient
\textit{D} is then obtained in the usual way: $D=\langle x^{2}\rangle/(2dt)$,
where $x$ is the defect displacement over the time $t$, and $d$ is the
dimension of the space. Typically, the residence times associated with moves
from visited sites are summed until the required time interval $t$ is
completed. But for purposes of this derivation, it is necessary to also regard
$t$ as the sum of the \textit{accrued} residence times at the visited sites.
This is because those accrued times, for sites in the basin, are proportional
to the equilibrium defect concentrations there. Also for purposes of the
derivation, it is convenient to use the term `periphery site'\ for those sites
in the basin from which the defect can move out of the basin. With this
terminology set, we obtain expressions for the probability $p_{i}$ of escape
from the basin via the particular periphery site $i$, and for the residence
time $t_{\text{basin}}$ associated with a visit by the defect to the basin.

Consider a defect at periphery site $i$. With each move from site $i$, the
accrued residence time for that site increases by an (average) amount
$\tau_{i}=\left(
%TCIMACRO{\tsum \nolimits_{b(i)}}%
%BeginExpansion
{\textstyle\sum\nolimits_{b(i)}}
%EndExpansion
k_{i\rightarrow b(i)}+%
%TCIMACRO{\tsum \nolimits_{q(i)}}%
%BeginExpansion
{\textstyle\sum\nolimits_{q(i)}}
%EndExpansion
k_{i\rightarrow q(i)}\right)  ^{-1}$, where the two sums are over all
transition rates from site $i$ to accessible sites $b(i)$ within the basin and
to accessible sites $q(i)$ outside the basin, respectively. The probability
that it will escape the basin on that move is $\epsilon_{i}=\left(
%TCIMACRO{\tsum \nolimits_{q(i)}}%
%BeginExpansion
{\textstyle\sum\nolimits_{q(i)}}
%EndExpansion
k_{i\rightarrow q(i)}\right)  \tau_{i}$; thus on average one of every
$\epsilon_{i}^{-1}$ visits by the defect to site $i$ will result in an escape.
In that event, the residence time $t_{i}^{e}\equiv\tau_{i}\epsilon_{i}%
^{-1}=\left(
%TCIMACRO{\tsum \nolimits_{q(i)}}%
%BeginExpansion
{\textstyle\sum\nolimits_{q(i)}}
%EndExpansion
k_{i\rightarrow q(i)}\right)  ^{-1}$, on average, has accrued to site $i$. It
is noteworthy that $t_{i}^{e}$ is a function only of the rates
$k_{i\rightarrow q(i)}$ out of the basin.

Of course, the basin may contain many periphery and interior sites. As the
defect moves within the basin, it produces an increasingly accurate set
$\{t_{k}/\langle t_{k}\rangle\}$ of \textit{relative} residence times, where
the sites $k$ are in the basin (periphery and interior) and the average
(indicated by the angle brackets) is taken over all sites in the basin. In
fact the elements $t_{k}/\langle t_{k}\rangle$ approach the values
$c_{k}/\langle c_{k}\rangle$, where $c_{k}$ is the equilibrium defect
concentration at site $k$ that is routinely obtained in molecular dynamics and
statics calculations ($c_{k}$ is an exponential function of the defect
formation energy at site $k$). During the time $T=%
%TCIMACRO{\tsum _{k}}%
%BeginExpansion
{\textstyle\sum_{k}}
%EndExpansion
t_{k}$, the number of visits by the defect to site $i$ is $t_{i}/\tau_{i}$.
Since the probability that a particular visit will \textit{not} lead to an
escape from the basin is $(1-\epsilon_{i})$, the \textit{a priori} probability
that the defect does \textit{not} escape from the basin via site $i$ during
the time interval $T$\ is $(1-\epsilon_{i})^{t_{i}/\tau_{i}}$. Then the
\textit{a priori} probability that the defect \textit{does} escape the basin
via site $i$ during the time interval $T$ is $1-(1-\epsilon_{i})^{t_{i}%
/\tau_{i}}\approx(t_{i}/\tau_{i})\epsilon_{i}$ for $\epsilon_{i}\ll1$. This
equals $t_{i}/t_{i}^{e}$ when the definition $t_{i}^{e}\equiv\tau_{i}%
\epsilon_{i}^{-1}$ is used. Thus the probability $p_{i}$ that the defect
escapes the basin from periphery site $i$ rather than from another periphery
site is given by%
\begin{equation}
p_{i}=(t_{i}/t_{i}^{e})\left(
%TCIMACRO{\tsum \limits_{j}}%
%BeginExpansion
{\textstyle\sum\limits_{j}}
%EndExpansion
(t_{j}/t_{j}^{e})\right)  ^{-1}\label{e1}%
\end{equation}
where the sum is over all periphery sites $j$. Substituting into Eq.
(\ref{e1}) the expression for $t_{i}^{e}$ gives%
\begin{equation}
p_{i}=t_{i}%
%TCIMACRO{\tsum \limits_{q(i)}}%
%BeginExpansion
{\textstyle\sum\limits_{q(i)}}
%EndExpansion
k_{i\rightarrow q(i)}\left[
%TCIMACRO{\tsum \limits_{j}}%
%BeginExpansion
{\textstyle\sum\limits_{j}}
%EndExpansion
\left(  t_{j}%
%TCIMACRO{\tsum \limits_{q(j)}}%
%BeginExpansion
{\textstyle\sum\limits_{q(j)}}
%EndExpansion
k_{j\rightarrow q(j)}\right)  \right]  ^{-1}\text{.}\label{e2}%
\end{equation}
As evident from this last equation, the escape from periphery site $i$ out of
the basin would be to site $q^{\prime}$ with probability $p_{i\rightarrow
q^{\prime}}=k_{i\rightarrow q^{\prime}}\left(
%TCIMACRO{\tsum \nolimits_{q(i)}}%
%BeginExpansion
{\textstyle\sum\nolimits_{q(i)}}
%EndExpansion
k_{i\rightarrow q(i)}\right)  ^{-1}$, where site $q^{\prime}$ is one of the
set $\{q(i)\}$. That is, a defect trapped in the basin will escape via
periphery site $i$ to site $q^{\prime}$ (outside the basin) with probability
$P_{i\rightarrow q^{\prime}}=p_{i}p_{i\rightarrow q^{\prime}}$.

The long-term, average behavior of the defect is thus reproduced by the
standard kMC method, with the addition that if the defect enters the basin, on
its next move it escapes the basin from a periphery site chosen in accordance
with the probability distribution implied by Eq. (\ref{e2}). The residence
time $t_{\text{basin}}$ associated with this move is given by the relation%
\begin{equation}
t_{\text{basin}}=%
%TCIMACRO{\tsum \limits_{j}}%
%BeginExpansion
{\textstyle\sum\limits_{j}}
%EndExpansion
t_{j}^{e}p_{j}+%
%TCIMACRO{\tsum \limits_{k}}%
%BeginExpansion
{\textstyle\sum\limits_{k}}
%EndExpansion
t_{k}\left(  t_{j}^{e}p_{j}/t_{j}\right)  \label{e3}%
\end{equation}
where the first sum is over all \textit{periphery} sites $j$, and the second
sum is over all basin \textit{interior} sites $k$. The second sum accounts for
the time the defect spends at interior sites, which in the case of a
particular interior site $k$ equals the ratio $t_{k}/t_{j}$ of time spent by
the defect at site $k$ to time spent at an arbitrarily chosen periphery site
$j$, multiplied by the average time $t_{j}^{e}p_{j}$ spent at site $j$ during
a visit by the defect to the basin (note that the ratio $t_{j}^{e}p_{j}/t_{j}$
is identical for all periphery sites $j$). The simplest example demonstrating
Eq. (\ref{e3}) is that of a basin comprised of $n$ \textit{identical}
periphery sites (that is, the probability $p_{j}$ of escaping the basin via a
particular periphery site is $1/n$, and all $t_{j}^{e}$ equal the
`lifetime'\ $t^{e}$) and no interior sites. Clearly the average residence time
in the basin per visit by the defect is $t^{e}$, so the average residence time
in each periphery site per visit to the basin must be $t^{e}/n$, which equals
$t_{j}^{e}p_{j}$ as expected.

By use of Eq. (\ref{e2}) for $p_{j}$, Eq. (\ref{e3}) may be rewritten as%
\begin{equation}
t_{\text{basin}}=%
%TCIMACRO{\tsum \limits_{k}}%
%BeginExpansion
{\textstyle\sum\limits_{k}}
%EndExpansion
t_{k}\left[
%TCIMACRO{\tsum \limits_{j}}%
%BeginExpansion
{\textstyle\sum\limits_{j}}
%EndExpansion
\left(  t_{j}%
%TCIMACRO{\tsum \limits_{q(j)}}%
%BeginExpansion
{\textstyle\sum\limits_{q(j)}}
%EndExpansion
k_{j\rightarrow q(j)}\right)  \right]  ^{-1}\label{e4}%
\end{equation}
where now the sum is over \textit{all} (periphery \textit{and} interior) basin
sites $k$. As discussed above, the ratio $t_{k}/t_{j}$ may be replaced by
$c_{k}/c_{j}$. Thus the equilibrating basin is accommodated by addition of the
set $\{P_{i\rightarrow q(i)}\}$ of probabilities for moves out of the basin,
and the residence time $t_{\text{basin}}$, to the kMC catalog of transition rates.

It may be noted that the derivation of Eq. (\ref{e1}) relies on the use of the
\textit{average} value (called $\tau$ above) for the time that accrues to a
basin site with each visit by the defect prior to escape. In conventional kMC
simulations, the time may alternatively be advanced by an amount $\triangle t$
taken randomly from the exponential distribution $\tau^{-1}\exp(-\triangle
t/\tau)$; that is, by the amount $\triangle t=\tau\lbrack-\ln z]$ where $z$ is
chosen randomly from the interval $(0,1]$. Thus it is possible in the latter
case to calculate the higher moments of the escape time from the basin as well
as the average time $t_{\text{basin}}$. Of course, the method developed here
for handling deep basins in kMC simulations presupposes that calculation of an
accurate distribution of basin escape times (whether desired or not) is not
computationally feasible. In this event, it is recommended (for consistency)
that \textit{average} values $\tau$, rather than variable values $\triangle
t$, be used to accrue time to sites outside the basin. This should not affect
the average value $\langle x^{2}\rangle$ obtained for a specified diffusion
time $t$, that is\ needed to calculate the defect diffusion coefficient $D$.

This method of handling a set of connected states may be contrasted with that
of Novotny \cite{r4}, who applies the finite Markov chain formalism \cite{r5}.
The basin sites are therefore \textit{transient} states, and the sites to
which the defect moves out of the basin are \textit{absorbing} states. All
transition probabilities connecting transient states, and connecting transient
states with absorbing states, are elements in the Markov transition matrix
$\mathbf{M}$. Then the formalism gives, for the defect in a specified initial
transient state, (1) the mean number of times in each of the transient states
before absorption, and (2) the probabilities for absorption in each of the
absorbing states. (See Ref. \cite{r6} for a detailed example of how to use
finite Markov chain theory to model stochastic physical systems.) The
correlation between the entrance and exit points at the basin periphery is
thus preserved at the expense of considerable mathematical and computational
complication (e.g., a different matrix $\mathbf{M}$ is needed for each of the
possible initial states). That virtue is minor when the defect is essentially
equilibrated in the basin before its escape, and in any event may be negated
by the various sources of error (e.g., inaccurate transition rates) and the
stochastic nature of the simulation. It should be emphasized that the Markov
approach requires that all transition rates between basin sites be available,
while the present approach can alternatively use equilibrium defect concentrations.

Before applying the method to sample systems with complex basins, it is
interesting to consider a very simple, one-dimensional system that can be
solved analytically. This is a linear arrangement of four sites, labeled (in
order) 1 through 4, where the transition rates $k_{2\rightarrow3}$ and
$k_{3\rightarrow2}$ are much faster than the rates $k_{2\rightarrow1}$ and
$k_{3\rightarrow4}$. Thus a defect will `flicker'\ between sites 2 and 3 many
times before escaping to site 1 or 4 \cite{r7}. The average behavior of the
defect in this system is easily calculated by use of the Markov formalism when
sites 1 and 4 are regarded as absorbing states. In the event that the defect
is initially at site 2, the analytic calculation produces the row vector%
\[
\mathbf{\beta}=\frac{1}{p_{2\rightarrow1}+p_{2\rightarrow3}p_{3\rightarrow4}}%
\begin{pmatrix}
p_{2\rightarrow1} & p_{2\rightarrow3}p_{3\rightarrow4} & 1 & p_{2\rightarrow3}%
\end{pmatrix}
\]
where $p_{i\rightarrow j}$ is the probability for the defect at site $i$ to
move to site $j$ [so, for example, $p_{2\rightarrow1}=k_{2\rightarrow
1}/(k_{2\rightarrow1}+k_{2\rightarrow3})$]; the elements $\beta_{1}$ and
$\beta_{2}$ are the probabilities for absorption at site 1 and site 4,
respectively; and the elements $\beta_{3}$ and $\beta_{4}$ are the mean number
of times at sites 2 and 3, respectively, before absorption. The expressions
for $\beta_{1}$ and $\beta_{2}$ have been obtained previously by Mason
\textit{et al}. \cite{r7}, by accounting for all possible numbers of flickers
prior to escape from sites 2 and 3: for example, the probability that a defect
initially at site 2 will escape to site 1 is $%
%TCIMACRO{\tsum \nolimits_{n=0}^{\infty}}%
%BeginExpansion
{\textstyle\sum\nolimits_{n=0}^{\infty}}
%EndExpansion
(p_{2\rightarrow3}p_{3\rightarrow2})^{n}p_{2\rightarrow1}=p_{2\rightarrow
1}/(1-p_{2\rightarrow3}p_{3\rightarrow2})$, which equals $\beta_{1}$.

In the event that the defect is initially at site 3, the corresponding
calculation produces the row vector%
\[
\mathbf{\beta}^{\prime}=\frac{1}{p_{2\rightarrow1}+p_{2\rightarrow
3}p_{3\rightarrow4}}%
\begin{pmatrix}
p_{3\rightarrow2}p_{2\rightarrow1} & p_{3\rightarrow4} & p_{3\rightarrow2} & 1
\end{pmatrix}
\text{.}%
\]
Then the `averaged'\ results are given by the row vector $\overline
{\mathbf{\beta}}=\chi_{2}\mathbf{\beta}+\chi_{3}\mathbf{\beta}^{\prime}$,
where $\chi_{2}$ and $\chi_{3}$ are relative concentrations at sites 2 and 3
that satisfy $\chi_{2}+\chi_{3}=1$ and detailed balance, $\chi_{2}%
k_{2\rightarrow3}=\chi_{3}k_{3\rightarrow2}$. Note that this averaging removes
any memory of the `initial'\ defect site (that is, whether the defect entered
from site 1 or from site 4). The averaged vector is%
\begin{align*}
\overline{\mathbf{\beta}} &  =\frac{1}{p_{2\rightarrow1}+p_{2\rightarrow
3}p_{3\rightarrow4}}\\
&
\begin{pmatrix}
\gamma_{1}p_{3\rightarrow2}p_{2\rightarrow1} & \gamma_{2}p_{2\rightarrow
3}p_{3\rightarrow4} & \gamma_{1}p_{3\rightarrow2} & \gamma_{2}p_{2\rightarrow
3}%
\end{pmatrix}
\end{align*}
where $\gamma_{1}=1+k_{3\rightarrow4}(k_{2\rightarrow3}+k_{3\rightarrow
2})^{-1}$ and $\gamma_{2}=1+k_{2\rightarrow1}(k_{2\rightarrow3}%
+k_{3\rightarrow2})^{-1}$. This may be compared with the equivalent row vector
$\mathbf{B}$ constructed from the stochastic quantities derived above for an
equilibrated basin:%
\begin{align*}
\mathbf{B} &  =%
\begin{pmatrix}
p_{2} & p_{3} & \frac{\chi_{2}t_{\text{basin}}}{\tau_{2}} & \frac{\chi
_{3}t_{\text{basin}}}{\tau_{3}}%
\end{pmatrix}
\\
&  =\frac{1}{p_{3\rightarrow2}p_{2\rightarrow1}+p_{2\rightarrow3}%
p_{3\rightarrow4}}\\
&
\begin{pmatrix}
p_{3\rightarrow2}p_{2\rightarrow1} & p_{2\rightarrow3}p_{3\rightarrow4} &
p_{3\rightarrow2} & p_{2\rightarrow3}%
\end{pmatrix}
\end{align*}
which very closely resembles $\overline{\mathbf{\beta}}$ when $p_{2\rightarrow
3}\gg p_{2\rightarrow1}$ and $p_{3\rightarrow2}\gg p_{3\rightarrow4}$.

A more complex basin is represented in Fig. 1.%
%TCIMACRO{\TeXButton{B}{\begin{figure}[tbp] \centering}}%
%BeginExpansion
\begin{figure}[tbp] \centering
%EndExpansion%
%TCIMACRO{\FRAME{itbpF}{3.0839in}{3.0891in}{0in}{}{}{vansiclen_fig1.eps}%
%{\special{ language "Scientific Word";  type "GRAPHIC";
%maintain-aspect-ratio TRUE;  display "USEDEF";  valid_file "F";
%width 3.0839in;  height 3.0891in;  depth 0in;  original-width 3.2059in;
%original-height 3.211in;  cropleft "0";  croptop "1";  cropright "1";
%cropbottom "0";  filename 'VanSiclen_fig1.eps';file-properties "XNPEU";}}}%
%BeginExpansion
{\includegraphics[
height=3.0891in,
width=3.0839in
]%
{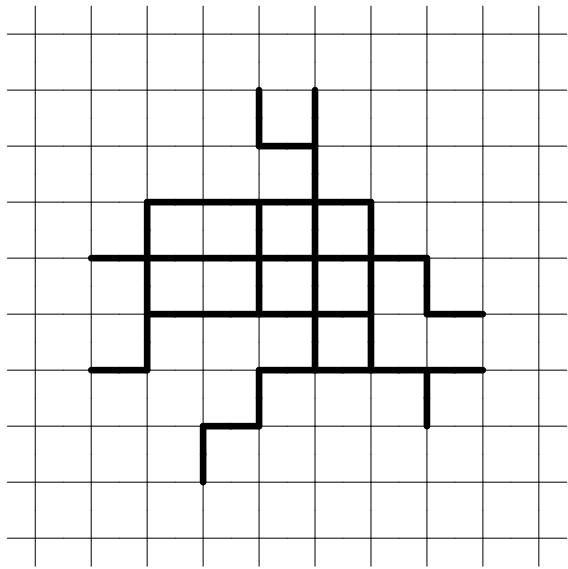}%
}%
%EndExpansion
\caption{Representation of a system of trapping and non-trapping sites. Those sites (nodes) connected by the thick bonds comprise the equilibrating basin in which the defect may be trapped for very long periods of time.\label{key}}%
%TCIMACRO{\TeXButton{E}{\end{figure}} }%
%BeginExpansion
\end{figure}
%EndExpansion
This system is a periodically repeated (in both dimensions) $10\times10$
regular network of nodes (defect-accessible sites) connected by bonds
(diffusion paths), where the `equilibrating basin'\ is the subset of 34 nodes
connected by the 40 thick bonds. Given the transition rates associated with
each bond, it is a straightforward matter to obtain the defect diffusion
coefficient by a kMC simulation.

Table I presents the diffusion coefficients calculated by the standard method
(`Exact') and by the `basin'\ method (`Approx.'), and an estimate of the
relative computation time needed in each case, for three different sets of
transition rates. The first set (row 1) has $k_{i\rightarrow j}=10\exp
[-(\mu_{i}-\mu_{j})]$ for the thick bonds and $k_{i\rightarrow j}=\exp
[-(\mu_{i}-\mu_{j})]$ for the thin bonds, where the $\{\mu_{i}\}$ are chemical
potentials assigned to the nodes with values taken randomly from the interval
$[0,1]$. The second set (row 2) is similar to the first set, but with the
difference that the $\mu_{i}$ for nodes belonging to the basin are taken from
the interval $[3,4]$, so that the defect will segregate to the basin. The
third set (row 3) is similar to the first set, but with the rates
$k_{i\rightarrow j}$ for the thick bonds having the prefactor 1000 (instead of
10). With these transition rates, detailed balance is satisfied:
$c_{i}k_{i\rightarrow j}=c_{j}k_{j\rightarrow i}$. The set $\{c_{i}\}$ is
needed to calculate the probabilities $\{p_{i}\}$ and the residence time
$t_{\text{basin}}$, and furthermore provides a nice check on the calculations
(namely, the accrued residence time $t_{i}$ at node $i$ should be proportional
to $c_{i}$). The values for the diffusion coefficient $D$ are believed to be
accurate to $\pm1$ in the last digit. In the last column, the `speed-up
factor'\ (due to use of the basin method) refers to the computational time
needed to accomplish a given defect diffusion time $t$, not to the
computational time needed to achieve a particular accuracy.%
%TCIMACRO{\TeXButton{B}{\begin{table*}[tbp] \centering}}%
%BeginExpansion
\begin{table*}[tbp] \centering
%EndExpansion%
\begin{tabular}
[c]{|c|c|c|c|}\hline
Transition rates & Exact $D$ & Approx. $D$ & Speed-up factor\\\hline%
\begin{tabular}
[c]{c}%
$k_{i\rightarrow j}^{\text{\textrm{(thick)}}}=10\exp[-(\mu_{i}-\mu_{j})]$\\
$k_{i\rightarrow j}^{\text{\textrm{(thin)}}}=\exp[-(\mu_{i}-\mu_{j})]$\\
$\mu_{i}\in\lbrack0,1]$%
\end{tabular}
& 1.373 & 1.799 & 1.2\\\hline%
\begin{tabular}
[c]{c}%
$k_{i\rightarrow j}^{\text{\textrm{(thick)}}}=10\exp[-(\mu_{i}-\mu_{j})]$\\
$k_{i\rightarrow j}^{\text{\textrm{(thin)}}}=\exp[-(\mu_{i}-\mu_{j})]$\\
$\mu_{i}^{\text{\textrm{(basin)}}}\in\lbrack3,4]$\\
$\mu_{i}^{\text{\textrm{(non-basin)}}}\in\lbrack0,1]$%
\end{tabular}
& 0.0285 & 0.0287 & 3.1\\\hline%
\begin{tabular}
[c]{c}%
$k_{i\rightarrow j}^{\text{\textrm{(thick)}}}=1000\exp[-(\mu_{i}-\mu_{j})]$\\
$k_{i\rightarrow j}^{\text{\textrm{(thin)}}}=\exp[-(\mu_{i}-\mu_{j})]$\\
$\mu_{i}\in\lbrack0,1]$%
\end{tabular}
& 1.79 & 1.799 & 81.0\\\hline
\end{tabular}
\caption{Comparison of diffusion coefficients calculated by the kinetic Monte Carlo method. The transition rates \emph{k} are for the paths represented by thick and thin bonds in Fig. 1; the $\mu$ are chemical potentials associated with the nodes that, for the purposes of this work, ensure that detailed balance is obeyed. The `Exact' and `Approx.' \emph{D} are diffusion coefficients calculated in the standard manner, and with the set of trapping states treated as an equilibrating basin, respectively. The `Speed-up factor' shows the computational advantage of the latter approach.\label{key}}%
%TCIMACRO{\TeXButton{E}{\end{table*}}}%
%BeginExpansion
\end{table*}%
%EndExpansion

The large difference in $D$ values in the first row of Table I shows that the
basin\ method does a poor job when the defect cannot equilibrate before
escaping; that is, when there is a significant spatial correlation between the
entry and exit points (in this case due to the small diffusivity contrast
between regions, which does not sufficiently confine the defect to the basin).
Otherwise, the diffusion coefficients obtained by assuming the defect to
equilibrate in the basin are seen to be very comparable to the
`exact'\ values, while costing (potentially) orders-of-magnitude less computer
time. Furthermore, the accrued residence times at the nodes (both inside and
outside the basin) are in every case extremely close to their exact values
(proportional to the $\{c_{i}\}$).

The results in Table I give a general indication of the utility of the basin
method. In particular, the method is accurate when the defect is essentially
equilibrated in the basin. The extent to which this is the case may be
determined by a conventional kMC simulation (where the basin method is not
used): the set $\{t_{k}/\langle t_{k}\rangle\}$ of relative residence times
for sites in the basin, obtained for a \textit{single} visit by the defect to
the basin, is compared with the set $\{c_{k}/\langle c_{k}\rangle\}$. The two
sets are more or less identical for a defect that is more or less equilibrated
in the basin.

In general the basin method gives an upper bound for the actual diffusion
coefficient. This is due to its neglect of any spatial correlation between the
entry and exit points at the basin periphery: the distance between these
points is, on average, less when they are spatially correlated than when they
are not. In either case the time spent in the basin per visit has average
value $t_{\text{basin}}$ (calculated according to the analytic expression
above), so a higher value for the diffusion coefficient is obtained in the
latter case. [That the average time spent in the basin per visit is
$t_{\text{basin}}$ in \textit{both} cases is evident from the fact that a kMC
simulation will produce a set $\{t_{m}/\langle t_{m}\rangle\}\approx
\{c_{m}/\langle c_{m}\rangle\}$ (where now \textit{all} sites $m$ in the
system---those outside the basin as well as those inside---are included),
whether the basin method is incorporated in the kMC code or not.] A comparison
of rows 1 and 3 in Table I illustrates this point. The two systems with
different sets of transition rates nonetheless possess (by design) identical
sets $\{c_{m}\}$, $\{p_{j}\}$, and $\{k_{j\rightarrow q(j)}\}$, and identical
basin residence time $t_{\text{basin}}$: this is the reason the two systems
produce the same `Approx.'\ value for the diffusion coefficient (1.799). But
the defect in the first system (row 1) is not well equilibrated in the basin,
causing an `Approx.'\ value for $D$ that is too high in that case.

As a final comment, it should be emphasized that this approach to
accommodating such trapping basins (created by, for example, segregation or
orders-of-magnitude differences in transition rates as considered in Table I)
in kMC simulations gives increasingly accurate results as the degree of
confinement increases, which is precisely the situation where kMC simulations
are, in the absence of this approach, increasingly inefficient and inaccurate.

This work was supported in part by the INL Laboratory Directed Research and
Development Program under DOE Idaho Operations Office Contract DE-AC07-05ID14517.

\end{document}